# Graphene-based field-effect transistor biosensors for the rapid detection and analysis of viruses: A perspective in view of COVID-19

*Joydip Sengupta[1] and Chaudhery Mustansar Hussain[2*]*

[1]Department of Electronic Science, Jogesh Chandra Chaudhuri College (Affiliated to University of Calcutta), Kolkata - 700 033, W.B., India

[2]Department of Chemistry and Environmental Science, New Jersey Institute of Technology, Newark, New Jersey, USA

**Abstract**

Current situation of COVID-19 demands a rapid, reliable, cost-effective, facile detection strategy to break the transmission chain and biosensor has emerged as a feasible solution for this purpose. Introduction of nanomaterials has undoubtedly improved the performance of biosensor and the addition of graphene enhanced the sensing ability to a peerless level. Amongst different graphene-based biosensing schemes, graphene field-effect transistor marked its unique presence owing to its ability of ultrasensitive and low-noise detection thereby facilitating instantaneous measurements even in the presence of small amounts of analytes. Recently, graphene field-effect transistor type biosensor is even successfully employed in rapid detection of SARS-CoV-2 and this triggers the interest of the scientific community in reviewing the current developments in graphene field-effect transistor. Subsequently, in this article, the recent progress in graphene field-effect transistor type biosensors for the detection of the virus is reviewed and challenges along with their strengths are discussed.

## 1. COVID-19 Transmission: Extent and Trivial Goals

In December 2019 a new strain of coronavirus caused severe respiratory illness[1] which was later termed as severe acute respiratory-related coronavirus 2 (SARS-CoV-2)[2]. World Health Organization (WHO) declared COVID-19 as pandemic on 13 March 2020 and urged international research community to carry out diagnostic test at a massive scale to fight the rapid transmission of the disease[3]. Since the transmission rate of SARS-CoV-2 is hasty fast, thus to break the chain of transmission a uniform diagnostic approach needs to be followed. WHO has formulated a guideline in this regard, named as ASSURED (Affordable, Sensitive, Specific, User-friendly, Rapid and robust, Equipment-free and Deliverable to end-users) as a benchmark to identify the most suitable diagnostic approach/test for resource-constrained situations[4]. However, this virus is novel in many aspects and possessed great difficulties for all the countries around the world to get control over it[5]. One of the major difficulties was that, some of the patients are asymptomatic but are capable of transmitting the virus[6]. Thus rapid, reliable testing needs to be developed for isolation of COVID-19 patients as well as for the safety for others.

## 2. Detection strategies for COVID-19

The early diagnosis of SARS-CoV-2 is crucial to prevent the severe outbreak of the disease and different detection strategies are being employed to encounter the adverse situation. Current detection strategies can be categorized as detection using immunological assays, detection based on amplification techniques and biosensing (Fig 1).

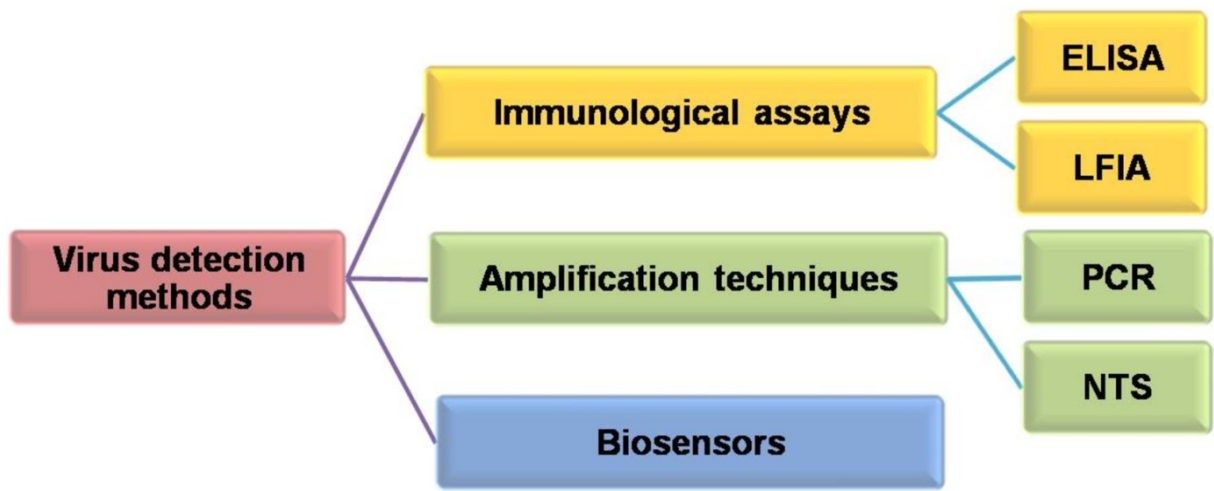

*Fig 1. Different virus detection methods.*

The methods involving immunological assays (e.g. Lateral flow immunoassay (LFIA), Enzyme-Linked Immunosorbent Assay (ELISA)) suffer from the drawback of the requirement of complex production routes of recombinant proteins and antibodies whereas amplification-based techniques (e.g. Reverse Transcription Polymerase Chain Reaction (RT-PCR), Nanopore Target Sequencing (NTS)) involve skilled personnel and expensive instruments[7]. Moreover, both of these methods require complex, expensive optical imaging devices and longer time for completion. Thus scientists are in search of an inexpensive, reliable, facile way for the apt detection of SARS-CoV-2 with high accuracy. Under the current scenario, biosensing is the most suitable technique for facile and accurate detection of the SARS-CoV-2. The biosensing is performed using a biosensor which is generally composed of two parts, a bioreceptor and a physiochemical transducer (Fig 2). The bioreceptor identifies the target analyte e.g. antibody, enzyme, DNA, aptamer etc. A biochemical signal is generated owing to the biomolecule-analyte interaction and transducer converts it into measurable information and analyzed quantitatively. The signal received from the transducer is firstly amplified and then processed for feeding into a display unit.

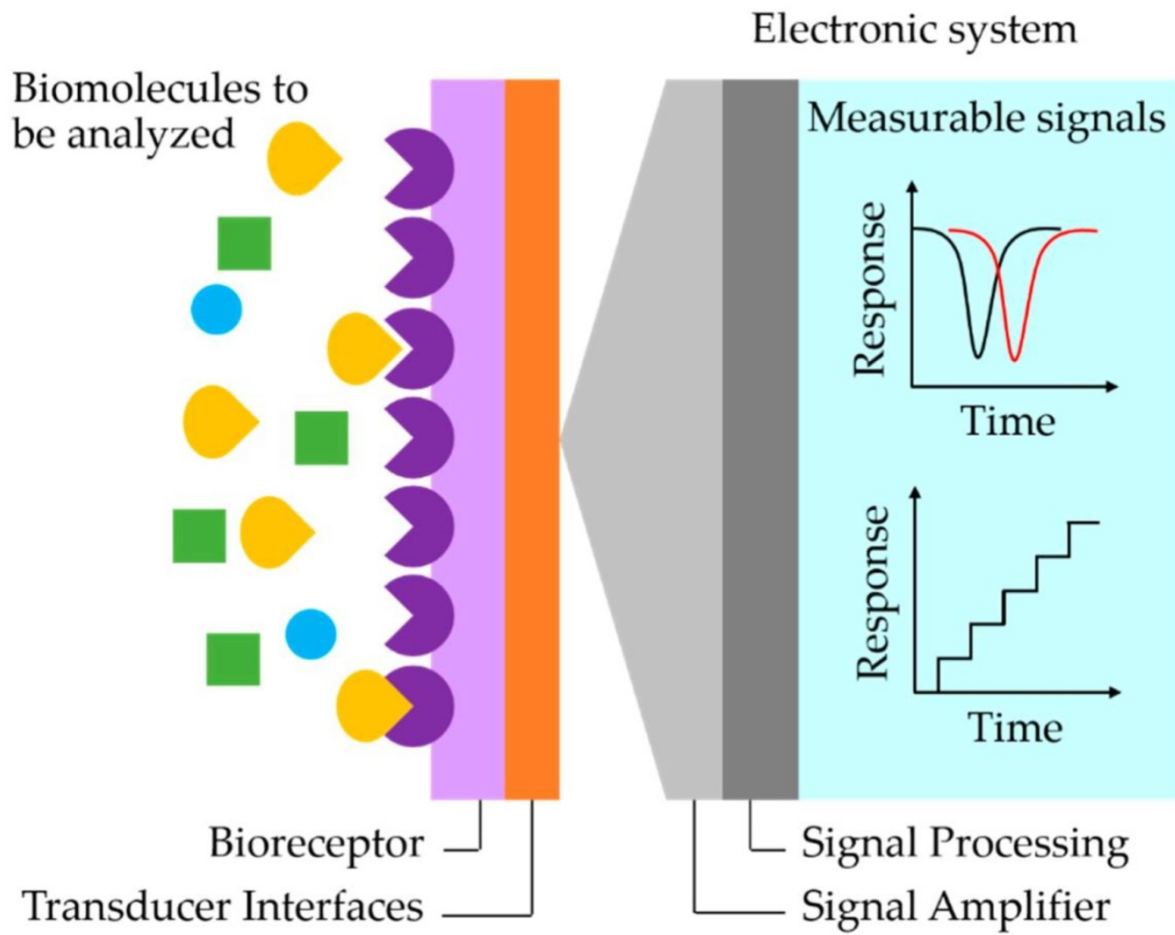

Fig 2. Schematic illustration of a typical biosensor system[8].

## 3. Nano-enabled biosensing systems for virus detection

Since the inception of the biosensor[9], a diverse range of biosensors undergoes research and development phase covering a broad spectrum of applications. However, a very limited number of biosensors are commercially available. For real-time sensing of biomarkers with picomolar (pM) sensitivity, only a few complex sensing systems involving high cost was proposed. Moreover, nanomaterials have been the area of keen interest since their discovery because of their unique and novel properties[10]. Consequently, various kinds (based on their size and shape) of nanomaterials are employed in biosensor for simple, reliable and inexpensive detection of viruses as listed in Table 1.

Table 1: Various nanomaterials used for biosensing of virus. Adapted from[11]

| **Nanomaterials** | **Key benefits** | **Reference** |
|---|---|---|
| Quantum dots | Excellent fluorescence, quantum confinement of charge carriers, and size-tunable band energy | Norouzi et al.[12] |
| Nanoparticles | Aid in immobilization, enable better loading of bio-analyte, and also possess good catalytic properties | Hamdy et al.[13] |
| Nanorods | Good plasmonic materials which can couple sensing phenomenon well and size-tunable energy regulation, can be coupled with MEMS, and induce specific field responses | Han et al.[14] |
| Nanowires | Highly versatile, good electrical and sensing properties for bio- and chemical sensing; charge conduction is better | Zhang et al.[15] |
| Carbon nanotubes | Improved enzyme loading, higher aspect ratios, ability to be functionalized, and better electrical communication | Bhattacharya et al.[16] |
| Graphene | High mechanical strength, Extreme conductivity, tunable bandgap, adjustable optical properties, large specific surface area | Pumera[17] |

Later on, graphene entered the biosensing-platform after the first demonstration of easy isolation of graphene by Geim and Novoselov, from bulk graphite in 2004[18]. Graphene is a two dimensional $sp^2$ hybridized form of carbon consisting of a single layer of graphite with a hexagonal lattice structure. Graphene can be oxidized by various routes to synthesize graphene oxide (GO) and the GO can be further made to undergo reduction schemes to achieve reduced GO (rGO) (Fig 3).

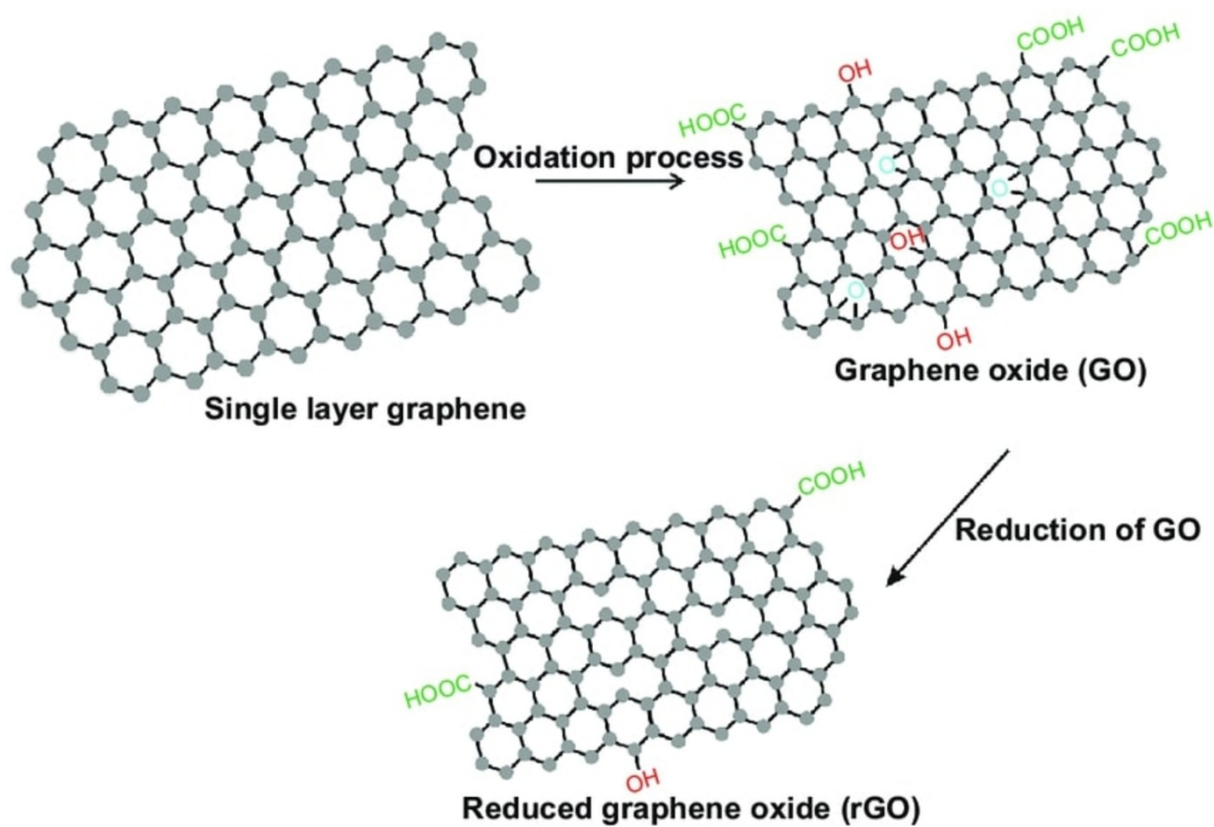

*Fig 3. Synthesis schemes of GO from graphene and rGO from GO[19].*

There are various methods to synthesize graphene and its derivatives for their integration in the biosensor; meanwhile, chemical vapour deposition[20] proves to be the best amongst them. Moreover, in recent days scientists have come up with advanced graphene synthesis method to curtail the production cost[21,22] even up to 95.5%[23]. In the pandemic scenario such as COVID-19, the cost of production of biosensing platforms is a major concern since millions of tests needs to be executed every day. The low cost of graphene makes it an ideal choice for the fabrication of high-performance and low-cost biosensing platforms.

Owing to the extraordinary physical and chemical properties, graphene-based materials have proved itself superior to other nanomaterials for the development of super-sensitive biosensors with a very low detection limit[24] (Fig. 4).

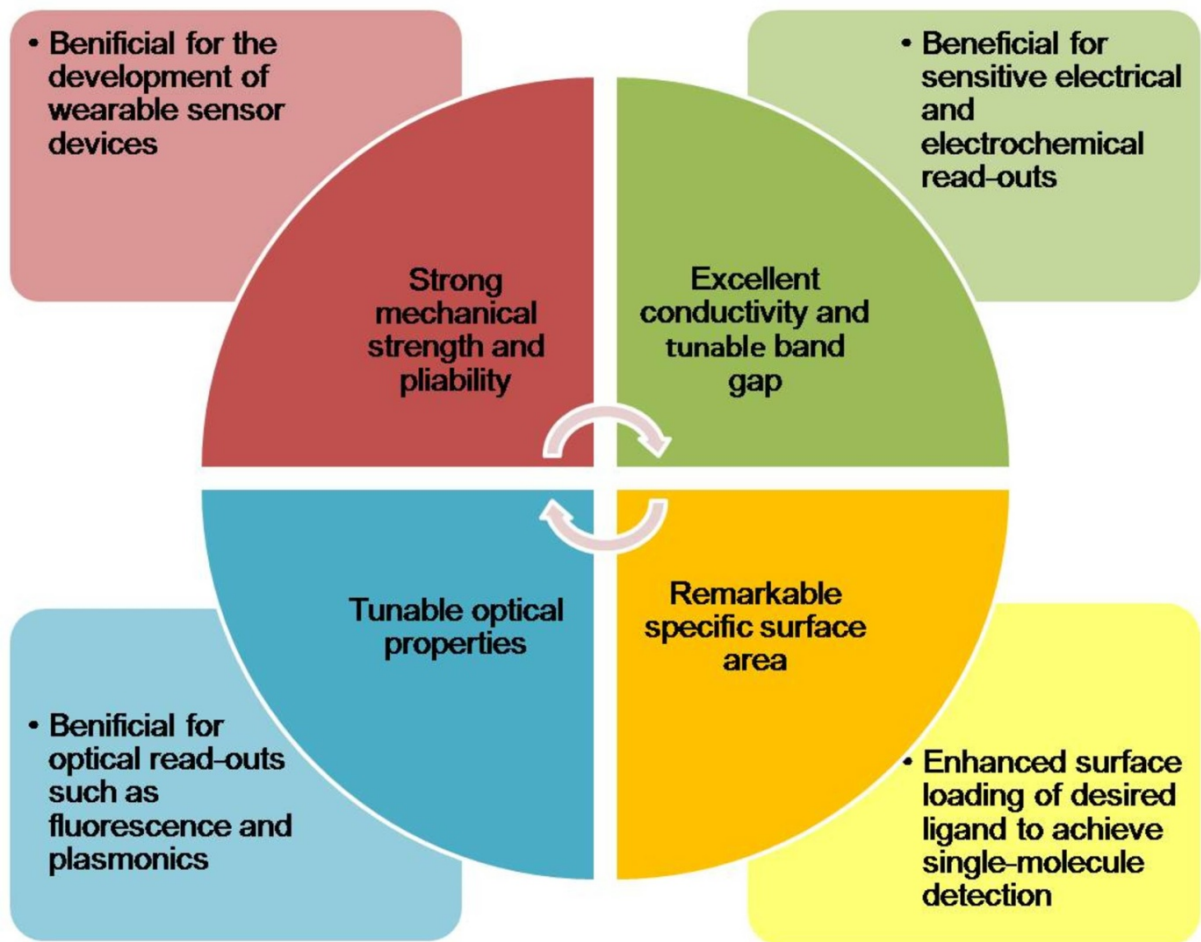

*Fig. 4. Extraordinary physical and chemical properties of graphene-based materials facilitate their use in biosensors.*

Subsequently, graphene-based sensing schemes have become widely popular nowadays[25] and are being employed by the researchers around the globe for detection of bacterial and viral pathogens[26], human virus[27], even for the vitamins[28] (e.g. biotin) with ultrahigh sensitivity and specificity.

## 4. Graphene-based electrochemical and optical biosensors for virus detection

### 4.1 Graphene-based electrochemical biosensors

Electrochemical impedance spectroscopy (EIS) makes use of the Faraday's law to characterize a chemical process employing electrical measurements. In EIS, a small oscillating voltage of varied frequencies is applied to an electrochemical cell and the electrical impedance signal is recorded to acquire information related to diffusion and charge transfer phenomena occurring on the surface of the electrode. An impedimetric biosensor is constructed employing this idea by immobilizing biological recognition elements onto the electrode surface of the electrochemical cell. Whenever the biological sample interacts with the biological recognition elements then the resulted biochemical activities is measured by the electrical impedance signal. Preferably the biorecognition layer should be deposited on a transducing material having a high conductivity, low electron transfer resistance and high surface to volume ratio. Since graphene qualifies for all the characteristics thus graphene-based impedimetric biosensor[29] was successfully employed for the detection of viruses. Glassy carbon electrode was modified by Gong et al.[30] using graphene-nafion composite film for the detection of human immunodeficiency virus (HIV) gene with a detection limit of 2.3 x $10^{-14}$ M. Muain et al.[31] used gold nanoparticles-decorated rGO nanocomposite for detection of hepatitis B virus with the lowest limit of detection of 3.80 ng mL$^{−1}$. For the detection of diarrhea virus, Li et al.[32] used nanocomposite consisting of molybdenum disulfide and rGO decorated with gold nanoparticles. However, there is a drawback of the impedimetric biosensor as the recording of the full impedance spectrum is a time-consuming task[33].

Impedimetric biosensor specifically measures the changes in charge conductance via computing the impedance of the electrode surface whereas electrochemical biosensor in general measures the biochemical interaction between a bioactive substance and a biomarker and in addition to measure the impedance (which measures sensitivity); it also selectively detects different molecules as the molecules can be oxidized/reduced at different potentials. Graphene is widely used in the electrochemical biosensor as the high surface area of graphene aids the presence of a large number of defects which act as electro-active sites for heterogeneous electron transfer[34]. Huang et al.[35] used silver nanoparticle-graphene-chitosan nanocomposite for the detection of avian influenza virus (H7) with a detection limit of 1.6pg/mL. Human influenza virus (H1N1) can also be detected using GO-based electrochemical biosensor as demonstrated by Joshi et al.[36] with a detection limit of 26 PFU/mL. Moreover, graphene is widely used for building analytical lab-on-chip platforms[37]. Sing et al.[38] explored this idea to fabricate rGO-based electrochemical immunosensor integrated with a microfluidic platform for the detection of influenza viruses with a lower detection limit of 0.5 PFU/mL. For the detection of dengue virus, Navakul et al.[39] used a graphene-polymer based electrochemical biosensor and the sensor could detect the virus down to the level of 0.12 PFU/mL. Graphene quantum dots[40] (GQD) were employed for the fabrication of electrochemical biosensor platform for the detection of hepatitis B and hepatitis E viruses by Xiang et al.[41] and Chowdhury et al.[42], respectively.

## 4.2 Graphene-based electrochemiluminescence and fluorescence biosensors

The novel optical properties of graphene such as broadband and tunable absorption along with remarkable polarization-dependent effects make it a suitable material for use in the optical biosensor. A material emits light when it undergoes different physical-chemical process such as electrochemiluminescence (emission of light as a result of a chemical reaction triggered by electron transfer process) and fluorescence (light emission caused by electronic relaxation). Graphene is applied in various roles in electrochemiluminescence[43] and fluorescence-based optical detection schemes. Wang et al.[44] used GO as fluorescence polarization signal amplifier for the detection of HIV with a detection limit of 38.6 pmol/L. GO is also employed for the detection of ebola virus by Wen et al.[45] via the exploration of quenching property of graphene. The lower detection limit of the device was 1.4pM. Jeong et al.[46] made a biosensor using GO and the fabricated biosensor could detect as low as 3.8 pg of influenza viral RNA. GQD is also widely used for optical biosensing of HIV[47] and dengue virus[48].

## 5. Graphene-based FET biosensors for virus detection

A field-effect transistor (FET) is a majority carrier device having three terminals namely source, drain and gate. In FET an electric field is applied at the gate terminal which in-effect modifies the conductivity of the channel placed between source and drain. In Graphene FET (GFET) graphene is introduced as a channel within the FET structure. Based on the fabrication scheme of the gate, the GFET can be categorized as back-gated GFET or top-gated GFET (Fig. 5 a and b). Since the conductivity of the graphene is very high thus the response time of GFET is extremely small which essentially aids the quick detection of the virus[49]. In GFET biosensing scheme the virus is immobilized on graphene surface and the virus modifies the conductivity of the graphene which is rapidly detected at the output (Fig 5 c and d).

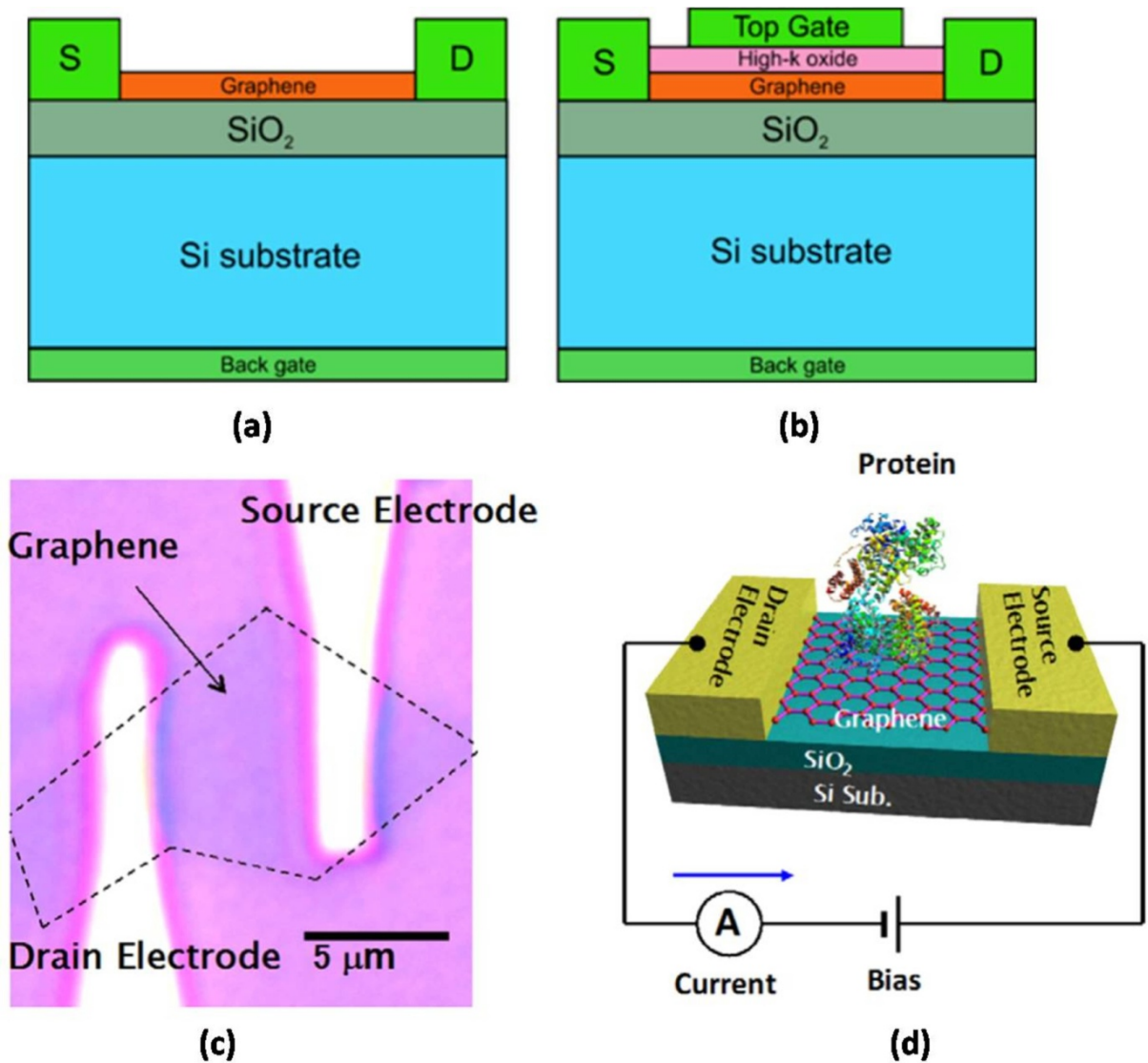

*Fig. 5. (a) Structure of back-gated GFET ; (b) Structure of top-gated GFET, (c) Optical microscope image of a GFET and (d) a schematic image of biosensing by the GFET.*[50,51]

Because of the immense potential of GFET in biosensing, scientists have used advanced modeling and simulation techniques to optimize the response of the GFET biosensor. Hamzah et al.[52] modeled the electrical transport of GFET biosensor and found that simulation temperature affects the electrical transport in GFET. Wu et al.[53] performed theoretical modeling of GFET biosensor for bacterial detection. They found that the probe size, graphene-bacteria distance and bacterial concentration are the key parameters for sensitive detection. Moreover, GFET biosensors often suffer from the issue of baseline drift in their response within the aqueous environment which makes it complicated to analyze biosensor response against target molecules. Ushiba et al.[54] employed state-space modeling to resolve this issue using time-series data of a GFET biosensor. Their model can be successfully employed for precise analysis of GFET biosensor response under aqueous environment. These theoretical modeling and simulation studies facilitate the optimization of the synthesis procedure of GFET biosensor to speed up the production.

Owing to scalability, miniaturization capability, rapid detection ability at low cost with high yield and reduced requirement of skilled personnel, GFET biosensors are widely used for the point-of-care (POC) diagnosis of different kind of viruses. Consequently, as the GFET biosensor needs to be used as a POC device, thus, it is of utmost importance to eliminate the possibility of cross-contamination mediated via the biosensor. Cartridge-type biosensor[55,56,57] is advantageous in this aspect, as the disposable cartridge eradicates the chances of cross-contamination and makes the readout interface reusable for numerous times (Fig 6).

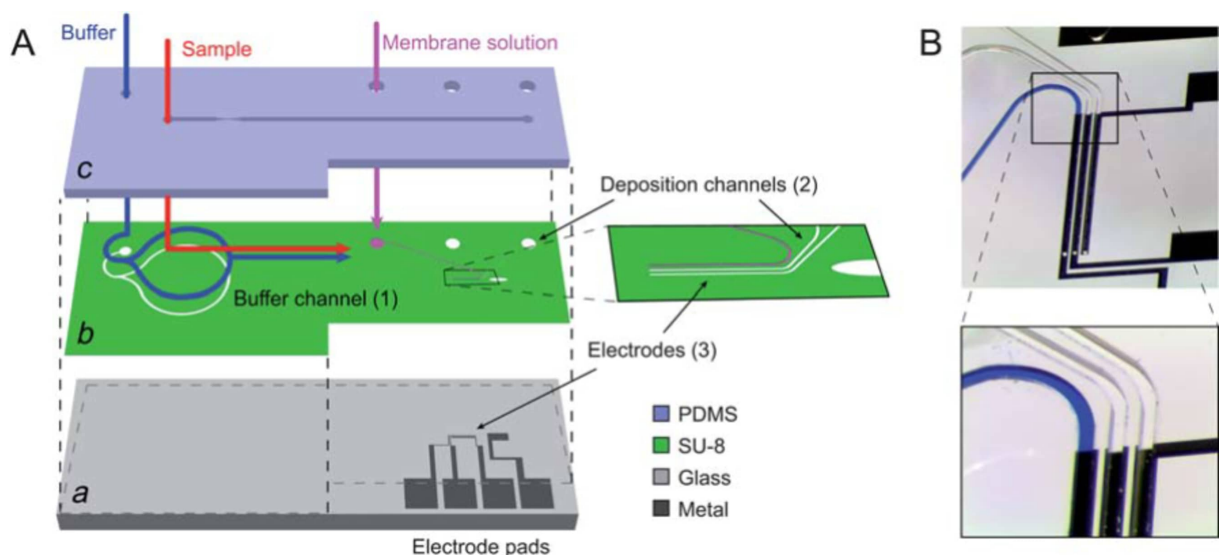

*Fig. 6. Design of a microfluidic cartridge used for biosensing[58].*

Chan et al.[59] adopted a flow-through approach to fabricate a microfluidic integrated rGO transistor-based DNA sensor for the detection of avian influenza virus. The strategy of extended long probe immobilization was proved to be beneficial for them as the minimum detection limit of 5 pM was achieved at a lesser time than electrochemical or optical sensing schemes. Graphene micro-electromechanical system (MEMS) technique was employed by Chen et al.[60] to fabricate a portable GFET biosensor for the detection of human influenza virus with a minimum detection limit of 1 ng/ml. Functionalisation is a useful tool for chemical modification of material surface structure and Ono et al.[61] used this option to modify the graphene surface for the fabrication of GFET biosensor. He used glycan for functionalisation and the fabricated GFET biosensor could successfully detect both the human and avian influenza virus with a detection limit of 130 pM and 600 nM, respectively. Mutsumoto et al.[62] used sugar chain to modify the GFET biosensor structure for the detection of both the human and avian influenza virus. He fabricated two kinds of GFET biosensor, one was modified with the human type sugar chain, another was modified by the bird type sugar chain and both types of GFET biosensor were successful in detecting the viruses. For the detection of ebola virus, Jin et al.[63] fabricated a FET structure using rGO as channel and after the immobilization of the virus on the surface; the response of the device was measured as a function of the Dirac voltage. The lower limit of detection of the device was 2.4 pg/mL. Chen et al.[64] also used the rGO based FET structure for the detection of ebola virus however the lower detection limit of the device was 1 ng/ml. Probabilistic neural network (PNN) was introduced by Ray et al.[65] for quantification of Hepatitis B virus using graphene nanogrids FET biosensor. The use of PNN proved to be around 85% better than polynomial fit and static neural network models in lowering of the detection limit. Using PNN the detection limit of the Hepatitis B virus could go down to 0.1 fM as reported by the group. Basu et al.[66] used nanoporous silicon oxide template to develop an rGO nanogrid based FET structure for the detection of Hepatitis B virus. Quantum transport behaviour of rGO and higher interaction of the biomolecule with the pore walls facilitate the sensitivity of the structure to achieve a detection limit of 50aM. Liquid ion-gated graphene nanohybrid FET structure decorated with close-packed carboxylated polypyrrole nano-particle arrays were synthesized by Kwon et al.[67] using the photolithographic approach. The scalable and flexible fluidic FET biosensor could sense the HIV at as low as 1 pM concentration. Islam et al.[68] used amine-functionalized graphene for the fabrication of FET structured biosensor. The fabricated biosensor was used for the detection of HIV and the results showed that that the biosensor could detect HIV with a limit of detection of 10 fg/mL. 3-D inkjet printing method was adopted by Xiang et al.[69] for the fabrication of GFET on flexible Kapton substrate for the detection of norovirus with a lower detection limit of 0.1 μg/ml. Aspermair et al.[70] fabricated aptamer-functionalized rGO-FET for the detection of human papillomavirus in saliva and the fabricated device could detect the virus with a detection limit of 1.75 nM. Hummers method along with photolithography and reduction process was employed by Liu et al.[71] for the fabrication micropatterned rGO FET. After the attachment of rotavirus on the graphene surface, the drain current quickly deceased and thus setting the lowest detection limit at $10^2$ pfu which is superior to the conventional ELISA method. Pant et al.[72] fabricated chemi-resistive rGO FET biosensor and used pyrene-NHS (Linker) complex for attachment of antibodies for the successful detection of rotavirus. For the detection of zika virus photolithography and plasma-enhanced chemical vapour deposition was employed by Afsahi et al.[73] to fabricate GFET on Si wafer. The inexpensive, portable GFET could detect the zika virus at concentrations as low as 450 pM. Currently, all the scientists around the globe are searching for a facile, reliable, inexpensive but sensitive detection scheme for the detection of SARS-CoV-2.

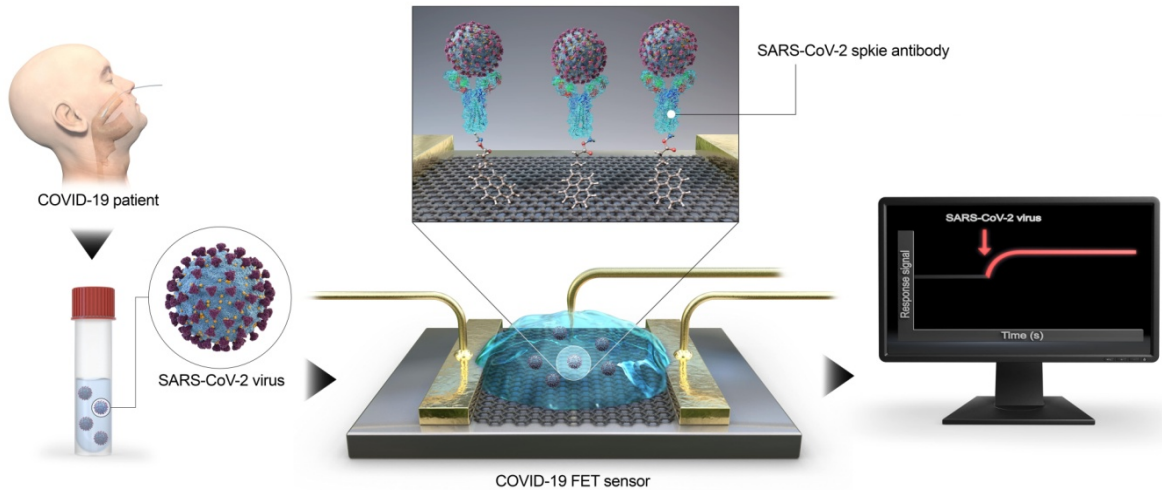

Fig. 7. SARS-CoV-2 detection scheme using GFET[74].

Seo et al.[74] fabricated a GFET functionalized with SARS-CoV-2 spike antibody for the detection of SARS-CoV-2 (Fig 7). 1-pyrenebutanoic acid succinimidyl ester was used as a probe linker for the immobilization of the SARS-CoV-2 spike antibody. The fabricated device could be used for detection of SARS-CoV-2 both in transport medium and clinical sample with a detection limit of 1 fg/ml. Zhang et al.[75] also fabricated GFET structure for the detection of SARS-CoV-2 within 2 mins with a minimum detection limit of 0.2 pM. The above discussion on the fabrication of GFETs and use them to detect various kind of virus is summarized in Table 2.

*Table 2. The overview of reported GFET biosensors for virus detection.*

| Virus Type | Basic Material | Limit of Detection | Reference |
|---|---|---|---|
| Avian influenza virus | rGO | 5 pM | Chan et al.[59] |
| Human influenza virus | Graphene | 1 ng/ml | Chen et al.[60] |
| Avian and Human influenza virus | Graphene | 130 pM (For human) 600 nM (For avian) | Ono et al.[61] |
| Avian and Human influenza virus | Graphene | NA* | Mutsumoto et al.[62] |
| Ebola Virus | rGO | 2.4 pg/mL | Jin et al.[63] |
| Ebola Virus | rGO | 1 ng/ml | Chen et al.[64] |
| Hepatitis B virus | Graphene | 0.1 fM | Ray et al.[65] |
| Hepatitis B virus | rGO | 50 aM | Basu et al.[66] |
| HIV | Graphene | 1 pM | Kwon et al.[67] |
| HIV | Graphene | 10 fg/mL | Islam et al.[68] |
| Norovirus | Graphene | 0.1 µg/ml | Xiang et al.[69] |
| Papillomavirus | rGO | 1.75 nM | Aspermair et al.[70] |
| Rotavirus | rGO | $10^2$ pfu | Liu et al.[71] |
| Rotavirus | rGO | NA* | Pant et al.[72] |
| Zika virus | Graphene | 450 pM | Afsahi et al.[73] |
| SARS-CoV-2 | Graphene | 1 fg/ml | Seo et al.[74] |
| SARS-CoV-2 | Graphene | 0.2 pM | Zhang et al.[75] |

* NA (data Not Available)

## 6. Challenges & drawback

Though biosensors have become an attractive choice for the detection of various kinds of the virus yet there are still some drawbacks exists. Graphene being a nanomaterial, to synthesize graphene with identical parameters is difficult and if any of the parameters varies then the response characteristic of the sensor may change. Sample preparation is a crucial step for reliable detection and it always been a hurdle for the scientist. Moreover, instead of the real virus environment often a dummy virus environment is used for the safety of the lab personnel. While many non-specific interactions may occur in the presence of the real sample and which might be of primary concern regarding the biosensor sensitivity issue. Controlling of binding efficiency between antibody and virus is also becomes difficult sometimes.

## 7. Outlook & Conclusions

Cost-effective and reproducible miniaturized GFET sensors having the potential for reliable diagnosis of the virus with high sensitivity and selectivity are required for the early-stage detection of SARS-CoV-2. Narvaez et al.[76] have made several recommendations in this regard among which integration of biosensor with the internet of things (IoT) is very important in terms of its use as a POC. To achieve these goals collaboration between physicists, chemists, material scientists, as well as engineers and medical personnel is of primary importance. Graphene is a wonder material as it showcased thorough several pieces of evidence, thus it can be forecasted that graphene will overcome all the barriers in future and GFET type biosensors will be the best possible biosensing scheme for graphene.

[44] Lijun Wang et al., 'A T7exonuclease-Assisted Target Recycling Amplification with Graphene Oxide Acting as the Signal Amplifier for Fluorescence Polarization Detection of Human Immunodeficiency Virus (HIV) DNA', *Luminescence* 31, no. 2 (2016): 573–79, https://doi.org/10.1002/bio.2997.

[45] Jia Wen et al., 'Study on Rolling Circle Amplification of Ebola Virus and Fluorescence Detection Based on Graphene Oxide', *Sensors and Actuators B: Chemical* 227 (1 May 2016): 655–59, https://doi.org/10.1016/j.snb.2016.01.036.

[46] Seonghwan Jeong et al., 'Fluorometric Detection of Influenza Viral RNA Using Graphene Oxide', *Analytical Biochemistry* 561–562 (15 November 2018): 66–69, https://doi.org/10.1016/j.ab.2018.09.015.

[47] Rong Sheng Li et al., 'Boron and Nitrogen Co-Doped Single-Layered Graphene Quantum Dots: A High-Affinity Platform for Visualizing the Dynamic Invasion of HIV DNA into Living Cells through Fluorescence Resonance Energy Transfer', *Journal of Materials Chemistry B* 5, no. 44 (15 November 2017): 8719–24, https://doi.org/10.1039/C7TB02356A.

[48] Nur Alia Sheh Omar et al., 'Sensitive Surface Plasmon Resonance Performance of Cadmium Sulfide Quantum Dots-Amine Functionalized Graphene Oxide Based Thin Film towards Dengue Virus E-Protein', *Optics & Laser Technology* 114 (1 June 2019): 204–8, https://doi.org/10.1016/j.optlastec.2019.01.038.

[49] Wang et al., 'Graphene Field-Effect Transistor Biosensor for Detection of Biotin with Ultrahigh Sensitivity and Specificity'.

[50] Kazuhiko Matsumoto et al., 'Recent Advances in Functional Graphene Biosensors', *Journal of Physics D: Applied Physics* 47, no. 9 (February 2014): 094005, https://doi.org/10.1088/0022-3727/47/9/094005.

[51] '4 Next Generation FETs Based on 2D Materials', accessed 8 September 2020, https://www.iue.tuwien.ac.at/phd/illarionov/dissch4.html.

[52] Azrul Azlan Hamzah, Reena S. Selvarajan, and Burhanuddin Y. Majlis, 'Electrical Characteristics of Graphene Based Field Effect Transistor (GFET) Biosensor for ADH Detection', in *Biosensing and Nanomedicine X*, ed. Hooman Mohseni, Massoud H. Agahi, and Manijeh Razeghi (Biosensing and Nanomedicine X, San Diego, United States: SPIE, 2017), 30, https://doi.org/10.1117/12.2273759.

[53] Guangfu Wu, M. Meyyappan, and King Wai Chiu Lai, 'Simulation of Graphene Field-Effect Transistor Biosensors for Bacterial Detection', *Sensors* 18, no. 6 (June 2018): 1715, https://doi.org/10.3390/s18061715.

[54] Shota Ushiba et al., 'State-Space Modeling for Dynamic Response of Graphene FET Biosensors', *Japanese Journal of Applied Physics* 59, no. SG (February 2020): SGGH04, https://doi.org/10.7567/1347-4065/ab65ac.

[55] Swee Ngin Tan et al., 'Paper-Based Enzyme Immobilization for Flow Injection Electrochemical Biosensor Integrated with Reagent-Loaded Cartridge toward Portable Modular Device', *Analytical Chemistry* 84, no. 22 (20 November 2012): 10071–76, https://doi.org/10.1021/ac302537r.

[56] Sascha Geidel et al., 'Integration of an Optical Ring Resonator Biosensor into a Self-Contained Microfluidic Cartridge with Active, Single-Shot Micropumps', *Micromachines* 7, no. 9 (September 2016): 153, https://doi.org/10.3390/mi7090153.

[57] M. Im et al., 'Development of a Point-of-Care Testing Platform With a Nanogap-Embedded Separated Double-Gate Field Effect Transistor Array and Its Readout System for Detection of Avian Influenza', *IEEE Sensors Journal* 11, no. 2 (February 2011): 351–60, https://doi.org/10.1109/JSEN.2010.2062502.

[58] Olivier Frey et al., 'Continuous-Flow Multi-Analyte Biosensor Cartridge with Controllable Linear Response Range', *Lab on a Chip* 10, no. 17 (7 September 2010): 2226–34, https://doi.org/10.1039/C004851H.

[59] Chunyu Chan et al., 'A Microfluidic Flow-through Chip Integrated with Reduced Graphene Oxide Transistor for Influenza Virus Gene Detection', *Sensors and Actuators B: Chemical* 251 (1 November 2017): 927–33, https://doi.org/10.1016/j.snb.2017.05.147.

[60] Yu-Jen Chen et al., 'Wireless Portable Graphene-FET Biosensor for Detecting H1N1 Virus', n.d., 1.

[61] Takao Ono et al., 'Glycan-Functionalized Graphene-FETs toward Selective Detection of Human-Infectious Avian Influenza Virus', *Japanese Journal of Applied Physics* 56, no. 3 (7 February 2017): 030302, https://doi.org/10.7567/JJAP.56.030302.

[62] Kazuhiko Matsumoto et al., 'Graphene Field-Effect Transistor for Biosensor', in *2016 23rd International Workshop on Active-Matrix Flatpanel Displays and Devices (AM-FPD)*, 2016, 45–46, https://doi.org/10.1109/AM-FPD.2016.7543613.

[63] Xin Jin et al., 'A Field Effect Transistor Modified with Reduced Graphene Oxide for Immunodetection of Ebola Virus', *Microchimica Acta* 186, no. 4 (7 March 2019): 223, https://doi.org/10.1007/s00604-019-3256-5.

[64] Yantao Chen et al., 'Field-Effect Transistor Biosensor for Rapid Detection of Ebola Antigen', *Scientific Reports* 7, no. 1 (8 September 2017): 10974, https://doi.org/10.1038/s41598-017-11387-7.

[65] R. Ray et al., 'Label-Free Biomolecule Detection in Physiological Solutions With Enhanced Sensitivity Using Graphene Nanogrids FET Biosensor', *IEEE Transactions on NanoBioscience* 17, no. 4 (October 2018): 433–42, https://doi.org/10.1109/TNB.2018.2863734.